\documentclass[prl,twocolumn,superscriptaddress,showpacs,preprintnumbers,amsmath,amssymb]{revtex4}
\usepackage{graphicx}
\usepackage{dcolumn}
\usepackage{bm}
\usepackage{mathrsfs}
\usepackage{subfigure}
\usepackage{natbib}

\begin{document}

\title{Realization of Artificial Ice Systems for Magnetic Vortices in a Superconducting MoGe Thin-film with Patterned Nanostructures}

\author{M. L. Latimer}
\affiliation{Materials Science Division, Argonne National Laboratory, Argonne, Illinois 60439, USA}
\affiliation{Department of Physics, Northern Illinois University, DeKalb, Illinois 60115, USA}
\author{G. R. Berdiyorov}
\affiliation{Departement Fysica, Universiteit Antwerpen, Groenenborgerlaan 171, B-2020 Antwerpen, Belgium}
\author{Z. L. Xiao}
\email{xiao@anl.gov}
\affiliation{Materials Science Division, Argonne National Laboratory, Argonne, Illinois 60439, USA}
\affiliation{Department of Physics, Northern Illinois University, DeKalb, Illinois 60115, USA}
\author{F. M. Peeters}
\email{francois.peeters@ua.ac.be}
\affiliation{Departement Fysica, Universiteit Antwerpen,
Groenenborgerlaan 171, B-2020 Antwerpen, Belgium}
\author{W. K. Kwok}
\affiliation{Materials Science Division, Argonne National Laboratory, Argonne, Illinois 60439, USA}
\date{\today}

\begin{abstract}

We report an anomalous matching effect in MoGe thin films containing pairs of circular holes arranged in such a way that four of those pairs meet at each vertex point of a square lattice. A remarkably pronounced fractional matching was observed in the magnetic field dependences of both the resistance and the critical current. At the half matching field the critical current can be even higher than that at zero field. This has never been observed before for vortices in superconductors with pinning arrays. Numerical simulations within the nonlinear Ginzburg-Landau theory reveal a square vortex ice configuration in the ground state at the half matching field and demonstrate similar characteristic features in the field dependence of the critical current, confirming the experimental realization of an artificial ice system for vortices for the first time.

\end{abstract}


\maketitle

Artificial ice systems \cite{Wang,Moller,Nisoli,Morgan,Mengotti,Ladak,Rougemaille,Budrikis,Brunner,Korda,Mangold,Babic,Han,Libal_col}
that can have properties similar to atomic spin ices \cite{Harris,Ramirez,Harris2,Lee,Hemberger,Mostovoy}
have been gaining tremendous interest in recent years in areas ranging from solid state systems, magnetism, and soft matter. Among them the two-dimensional (2D) artificial spin ices created using e.g., nanomagnetic arrays \cite{Wang,Moller,Nisoli,Morgan,Mengotti,Ladak,Rougemaille,Budrikis} and charged colloidal particle assemblies \cite{Brunner,Korda,Mangold,Babic,Han,Libal_col} have opened a new avenue in the study of novel phenomena such as geometrical frustration \cite{Rougemaille,Budrikis,Harris,Ramirez,Harris2,Lee,Hemberger,Mostovoy,Laughlin,Castelnovo,Choudhury,Anderson} which can elucidate, e.g., exotic spin states,\cite{Ramirez} charge quantization in magnetic monopoles,\cite{Laughlin,Castelnovo} and mechanisms of high-$T_c$ superconductivity.\cite{Anderson} In artificial nanomagnetic square spin ices,\cite{Wang,Budrikis} however, the ice rule states with spin arrangements following "two spins in, two spins out" orders \cite{Harris2} have been only partially observed, which could be due to the weak interactions between the magnetic islands.

Vortex matter in a superconductor has much stronger interactions relative to the pinning strength due to the much smaller size scale of the pinning array and could therefore permit a true ice rule obeying ground state. In a recent theoretical work Libal {\it et al.} proposed to create artificial square and Kagome ices with vortices in superconductors containing nanostructured arrays of pinning centers.\cite{libal} Using elongated double-well pinning sites arranged in a square lattice, for example, they were able to obtain the ground state of a square vortex ice which follows the "two vortices in, two vortices out" rule at each vertex, where the state of each double-well site is defined as "in" if the vortex sits close to the vertex and "out" otherwise. Such vortex systems can offer several advantages over the other artificial ices:\cite{libal} i) the ground state can be reached more rapidly as compared to nanomagnet systems due to the larger vortex-vortex interaction strength; ii) defect formation processes can be studied by changing the magnetic field to create vacancies or interstitials that locally break the ice rules; iii) different dynamical annealing protocols can be realized by an applied drive; and iv) transport properties of the system can be measured, which are not accessible in other ice systems.

\begin{figure}[t]
\includegraphics[width=\linewidth]{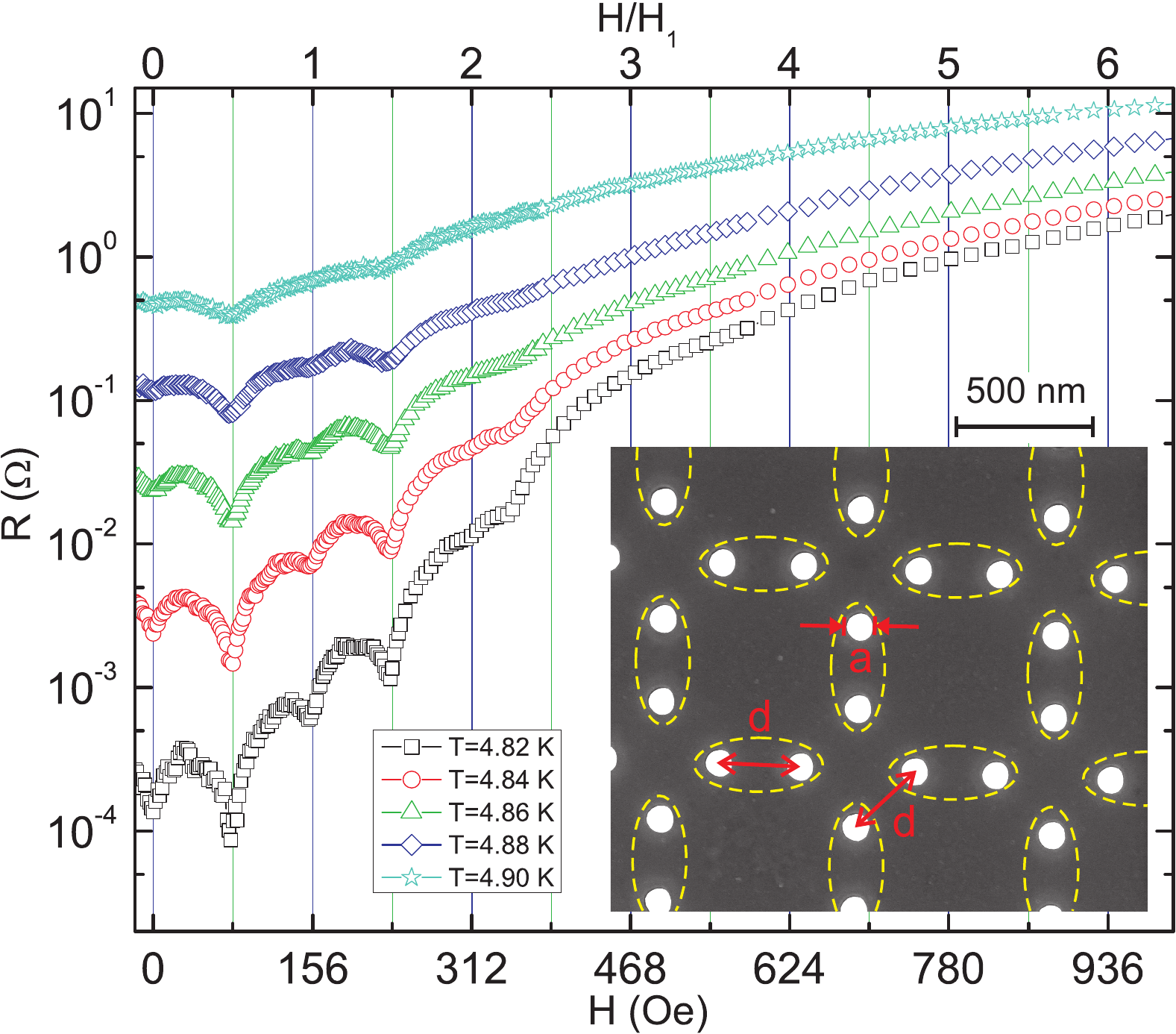}
\caption{\label{fig1}(color online) Measured resistance vs. magnetic field curves for a 20-nm-thick MoGe film with a square ice arrangement of holes (diameter $a=102$ nm and inter-hole spacing $d=300$ nm) at different temperatures for the applied dc current $I=500~\mu$A. Inset shows a scanning electron microscopy image of the sample. Dashed ellipses highlight the pair of holes (which are highlighted in white) that creates an effective double-well pinning potential for the vortices.}
\end{figure}

Here we report the first attempt to experimentally achieve vortex ice in superconducting systems. This work not only confirms previous theoretical prediction on the vortex ice ground state \cite{libal} but also reveals novel transport phenomena that have not been seen in both experiment and theory. We fabricated superconducting thin films containing pairs of circular holes arranged in such a way that four of those pairs meet at each vertex point of a square lattice (see inset of Fig. \ref{fig1}). Each pair of holes imitates the elongated double-well pinning sites considered in theory.\cite{libal} We measured the magnetic field dependences of the critical current and the magnetoresistance of the patterned films at various temperatures and found unusual matching effect which is different from that reported before: \cite{Moshchalkov,Metlushko,penrose,MoGe} fractional matchings are more pronounced than those at the usual integer matching fields. In particular, at half matching field the magnetoresistance can be smaller and the critical current can become larger than those at zero magnetic field. Numerical simulations within the time-dependent Ginzburg-Landau (GL) theory, where the vortex-vortex and vortex-hole interactions are accurately taken into account, attribute the observed unique characteristics to the square-ice arrangement of vortices, revealing signatures of square vortex ice in transport measurements.

Experiments were carried out on MoGe thin films which are known for their weak random vortex pinning, enabling transport measurements in a large temperature range.\cite{zhili} 20 nm thick films were sputtered from a Mo$_{0.79}$Ge$_{0.21}$ alloy target onto a silicon substrate with 200-nm-thick oxide layer separation. The samples were first patterned into a 50 $\mu$m wide microbridge using photolithography. Pairs of holes with various diameters $a$ and hole-hole spacings $d$ arranged in a square lattice (see inset of Fig. \ref{fig1}) were introduced into the section between the two voltage leads which are 50 $\mu$m apart, using focused-ion-beam milling (FEI Nova 600, 30 KeV Ga$^+$, 10-20 nm beam diameter). Transport measurements were carried out using a standard dc four-probe method with a Physical Property Measurement System (PPMS-9, Quantum Design, Inc). We measured three samples with hole-hole spacings of $d$=200 nm, 300 nm and 400 nm. Here we present the results for the sample with $d=300$ nm and $a=102$ nm. The resistance criterion 0.5$R_n$ (with $R_n$ the normal state resistance) gives us the zero-field critical temperature $T_{c0}$=5.1 K. Zero temperature coherence length $\xi(0)$ and the penetration depth $\lambda(0)$ were estimated to be equal to 6 nm and 400 nm, respectively.\cite{MoGe}

\begin{figure}[b]
\includegraphics[width=\linewidth]{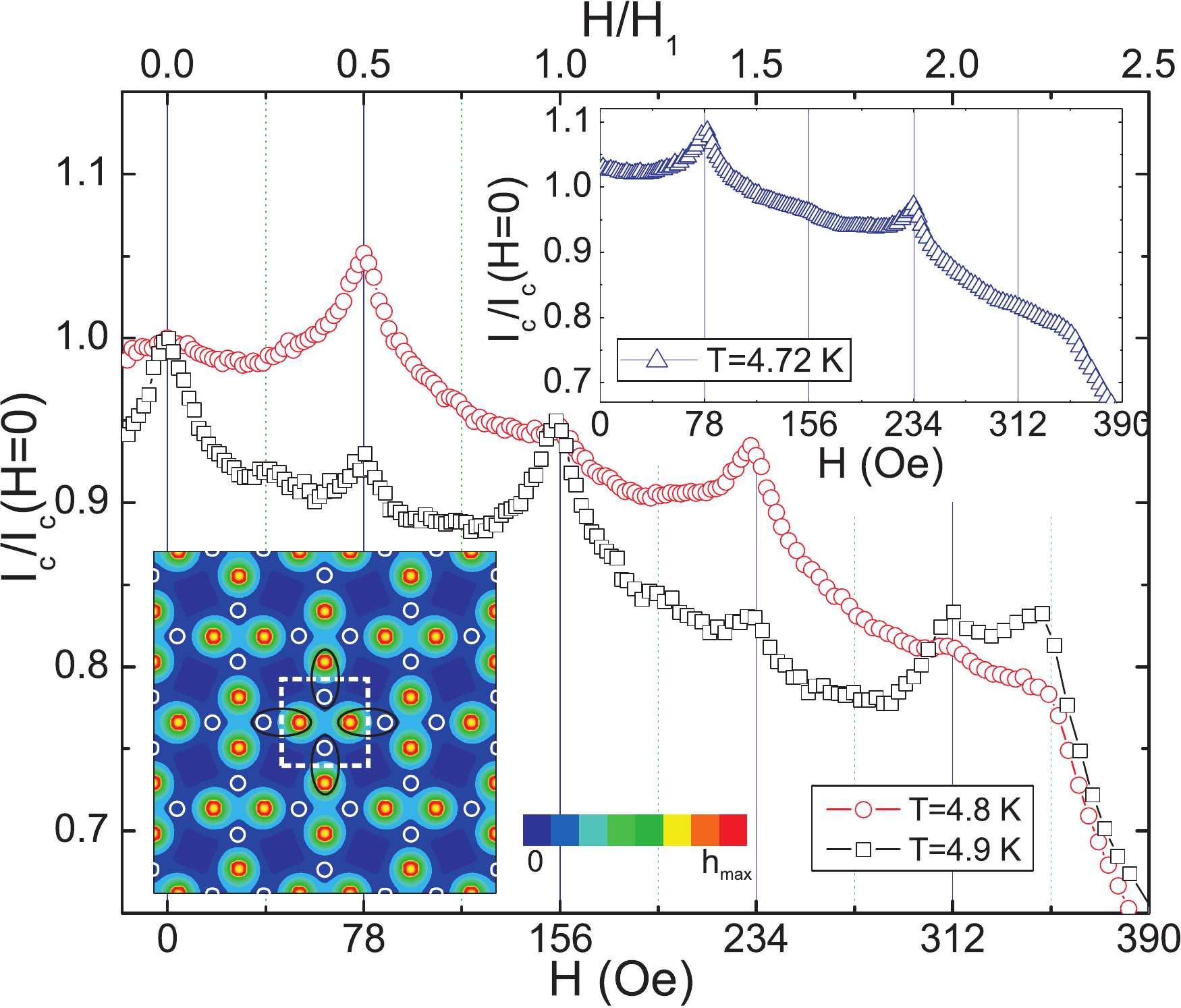}
\caption{\label{fig2}(color online) The critical current of the sample in Fig. \ref{fig1} (normalized to zero magnetic field critical current $I_c(0)$) as a function of the magnetic field $H$ at $T=4.8$ K (red circles) and $T=4.9$ K (black squares). Top inset shows the results obtained at $T=4.72$ K. Zero magnetic field critical currents are: $I_c(0)=802~\mu$A at $T=4.72$ K, $I_c(0)=613~\mu$A at $T=4.8$ K and $I_c(0)=335~\mu$A at $T=4.9$ K. Lower inset shows contour plots of the simulated local magnetic field $h$ for the half matching field, which shows the ground state vortex configuration at $T=4.8$ K. White circles indicate the positions of vortex-free holes and the dotted white square shows the unit cell of the simulation area.}
\end{figure}

Figure \ref{fig1} shows measured magnetoresistance curves of the sample at different temperatures. As expected, resistance dip induced by commensurate pinning enhancement is observed at the first matching field $H_1=156$ Oe when each hole pins one flux quantum (see the top axis in Fig. \ref{fig1}). However, a more pronounced dip in the $R(H)$ curve is found at the half matching field. Remarkably, the resistances at half matching field are smaller than the values at zero field for all temperatures considered. To our knowledge, this has not been reported up to now for other arrangements of artificial pinning centers \cite{Moshchalkov,Metlushko,penrose,MoGe}. The amplitude of the resistance dips increases with decreasing temperatures. At low temperatures (e.g. $T$=4.82 K) fractional matching can be even identified at 9/4$H_1$, indicating the presence of caged interstitial vortices, i.e. one interstitial vortex in the center of each unit cell caged by two vortices sit in each hole.\cite{MoGe,BerdiyorovEPL}

Figure \ref{fig2} shows the magnetic field dependence of the critical current of the sample at different temperatures, which is obtained from a series of $I-V$ curves with the voltage criterion $V^*=1~\mu$V. These $I_c(H)$ curves exhibit similar commensurate pinning enhancement features of the $R(H)$ curves at both fractional and integer matching fields as presented in Fig. \ref{fig1}. A remarkable new feature is that the critical current at half matching field can be larger than that at zero-field at low temperatures (see $I_c(H)$ curves at 4.72 K and 4.8 K). This is contrary to conventional wisdom, considering the fact that moving vortices are the main source of energy dissipation in a type-II superconductor and the critical currents decrease with increasing magnetic field.

To provide better insight into the vortex arrangements in the system we conducted simulations within the GL formalism, following the numerical approach of Refs. \cite{golibPRL,golibPRB}. This approach has been successfully used in ``visualizing'' different vortex configurations in perforated superconductors exhibiting close agreement with experiment.\cite{golibPRB} Our model system consists of a square simulation region with $4\times4$ unit cells, which is exposed to a homogeneous magnetic field $H$. Each unit cell (indicated by the white dotted square in the lower inset of Fig. \ref{fig2}) represents a meeting point for four holes in neighboring effective double-pinning wells (highlighted by black ellipses). Periodic boundary conditions are used in all directions in the 2D plane. For a given magnetic field, the ground state vortex configuration is obtained in ``field-cooled'' simulations starting from different random initial conditions. The lower inset in Fig. \ref{fig2} shows the lowest energy configuration for $H=0.5H_1$ at $T=4.8$ K. Vortices form an ordered structure -- they sit in one of the effective pinning sites, alternating their position from one hole to the other, due to the strong interaction with the neighboring vortices. In other words, the system obeys the ``ice rules'' of ``two vortices in, two vortices out'' around each vertex point, as predicted in Ref. \cite{libal}. The same vortex-ice structure is found in the ground state for the other temperatures considered. Since vortex arrangement is well correlated with the observed critical current of the sample (see, e.g., Ref. \cite{MoGe}), the critical current enhancement in such systems is a clear manifestation of the vortex ice formation process.

\begin{figure}[t]
\includegraphics[width=\linewidth]{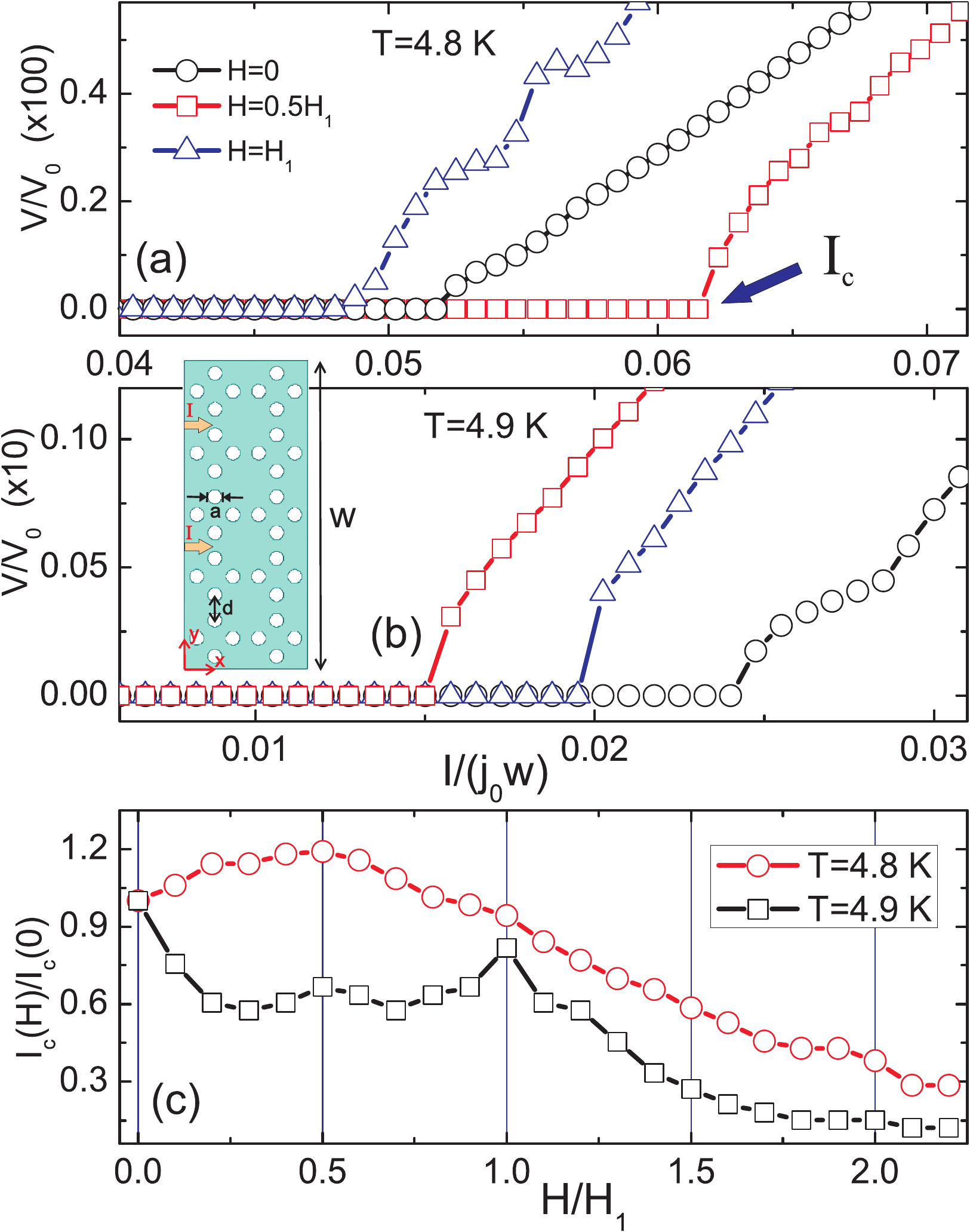}
\caption{\label{fig3}(color online) Simulation results. (a,b) Current-voltage characteristics of the sample with $a=102$ nm and $d=300$ nm at $T=4.8$ K (a) and $T=4.9$ K (b) for applied field values: $H=0$ (circles), $H=0.5H_1$ (squares) and $H=H_1$ (triangles). Inset in (b) shows the model system: a superconducting strip (width $w$ and with periodic boundary condition in the $x$-direction) with a square ice arrangement of holes (size $a$ and separation $d$) under applied dc current $I$ and a perpendicular magnetic field $H$. (c) The critical current of the sample (normalized to the one at zero magnetic field) as a function of $H$ at two temperatures.}
\end{figure}

In what follows, we conduct numerical simulations within the time-dependent GL theory to study the dynamics of superconducting vortices in the model geometry shown in the inset of Fig. \ref{fig3}(b). We solve the following set of equations \cite{Kramer}:
\begin{eqnarray}
u\frac{\partial\psi}{\partial t}=(\bigtriangledown-i\mathbf{A})^2\psi+\left(1-T-|\psi|^2\right)\psi,\\
\frac{\partial {\bf A}}{\partial
t}=\textrm{Re}\left[\psi^*(-i\nabla-{\bf A})\psi\right]-\kappa^2
\textrm{rot rot} {\bf A},
\end{eqnarray}
with units: the coherence length $\xi(0)$ for the distance, $t_{GL}(0)$$=$$\pi \hbar \big/8k_BT_cu$ for time, $T_c(0)$ for
temperature, and $\phi_0/2\pi\xi(0)$ for the vector potential ${\bf A}$. The current density is in units of $j_0=\sigma_n\hbar/2et_{GL}(0)\xi(0)$ and the voltage is measured in $V_0=\phi_0/2\pi t_{GL}$. The relaxation parameter $u$ is chosen as $u=1$ which results in optimal agreement with the experiment (see, e.g., Ref. \cite{golibPRL12}). The above equations are solved self-consistently using the numerical approach of Ref. \cite{golibPRL12} with periodic boundary conditions in the $x$-direction and Neumann boundary conditions at all sample edges. The results are obtained for $\kappa=10$.

As a representative example, we consider a superconducting strip with 2$\times$5 unit cells in the presence of transport current and perpendicular magnetic field (see inset of Fig. \ref{fig3}(b)). The inter-hole spacing and the size of the holes are the same as in the experiment (i.e., $d=300$ nm and $a=102$ nm). Figures \ref{fig3}(a,b) show the $I-V$ characteristics of the sample for three values of the applied magnetic field. It is seen from this figure that at lower temperature (Fig. \ref{fig3}(a)) the threshold depinning current of the sample ($I_c$, first jump in the $I-V$ curves) increases with magnetic field (compare circles and squares). This critical current decreases again with further increasing the applied field (triangles). However, such a behavior in the critical current is not found at higher temperatures (Fig. \ref{fig3}(b)). Figure \ref{fig3}(c) summarizes our simulation results on the critical current $I_c$ of the sample as a function of applied magnetic field. The figure shows that at low temperature (red circles) $I_c$ increases with applying small magnetic field and reaches its maximum at the half matching field. No peak is observed near the first matching field. At higher temperature (black squares) the $I_c$ enhancement near half matching field is strongly suppressed and a peak in the $I_c(H)$ curve appears near the first matching field. That is, the critical current has a similar dependence on magnetic field and temperature, as found in our experiment (see Fig. \ref{fig2}).

\begin{figure}[b]
\includegraphics[width=\linewidth]{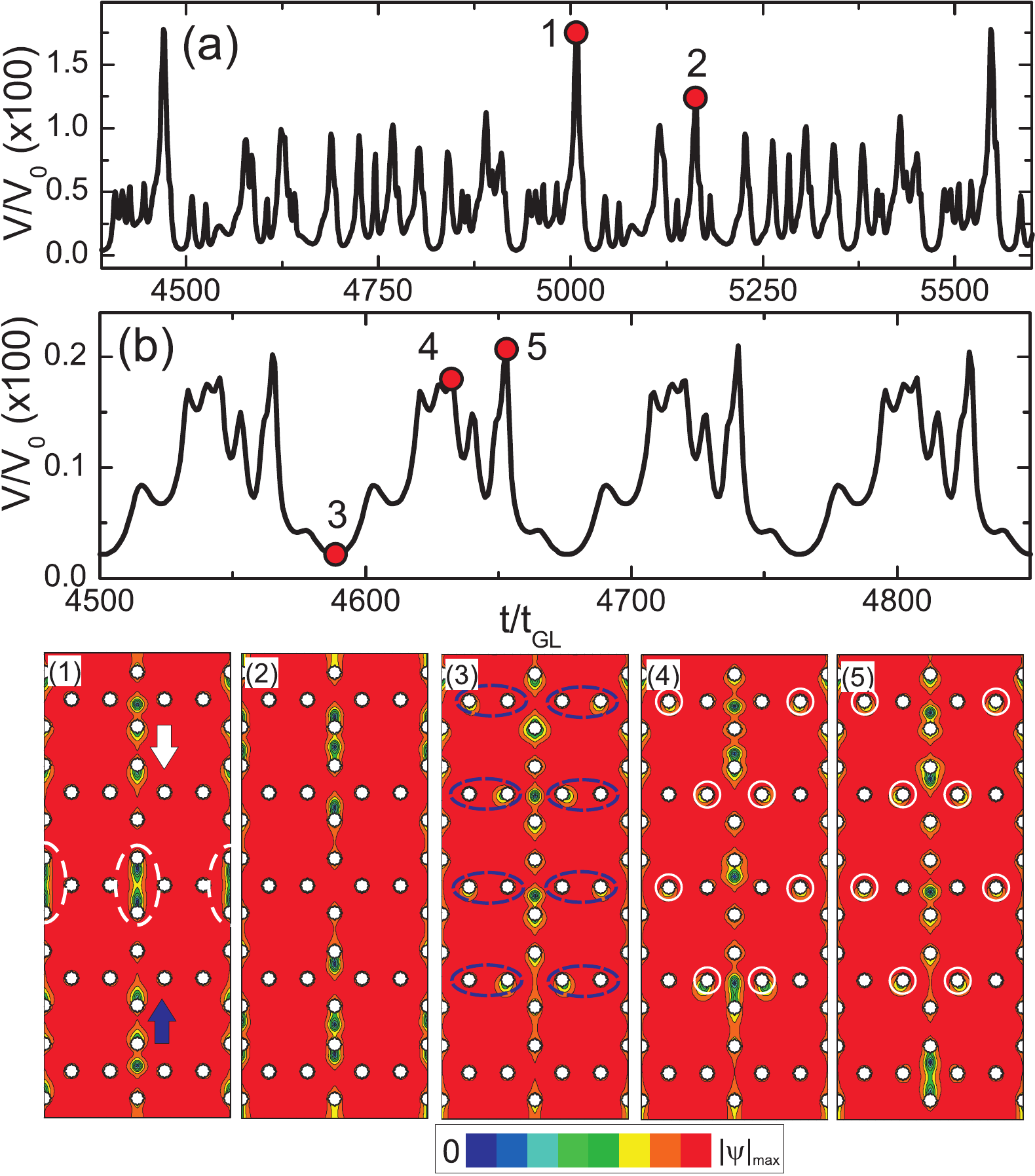}
\caption{\label{fig4}(color online) (a,b) Simulated voltage vs time response of the sample of Fig. \ref{fig3} at $T=4.8$ K for zero (a) and half matching field (b). Panels 1-5 show snapshots of $|\psi|^2$ at times indicated on the $V(t)$ curve. White/blue arrow in panel 1 shows the direction of motion of vortices/antivortices and dashed white ellipses highlight the annihilation point of vortex-antivortex pairs. Blue ellipses in panel 3 show the square ice arrangement of vortices in neighboring unit cells. White circles in panels 4 and 5 show positions of holes with a pinned vortex.}
\end{figure}

To understand the mechanism for such a field-enhanced critical current phenomenon near half matching field, we plot in Figs. \ref{fig4}(a,b) the simulated time evolution of the output voltage at $H=0$ (a) and $H=0.5H_{1}$ (b) for applied current $I/w=0.063j_0$, together with snapshots of the Cooper-pair density at times indicated on the $V(t)$ curves. At zero field, the dissipation is due to the motion of vortices and antivortices, which nucleate at the opposite outer boundaries and move towards the middle of the sample where they annihilate (panel 1). Their motion from one hole to another, as well as their annihilation gives voltage oscillations across the sample (panel 2, see also supplemental online video showing the motion of vortices \cite{suppl2}).

As we have already shown in the inset of Fig. \ref{fig2}, the ground state configuration of the system at half matching field for the given parameters consists of the vortex ice structure. At small current values the system preserves the ice rule which is highlighted by the blue dashed ellipses in panel 3 of Fig. \ref{fig4}. As in the case of zero magnetic field, vortices predominantly move along a straight line connecting the closest pinning centers. The pinned vortices near such a channel (highlighted by white circles in panels 4 and 5) introduce an extra potential for the moving vortices.\cite{BerdiyorovEPL} Since such pinned vortices near the sample edge prevent the penetration of additional vortices into the system, we observed the motion of vortices in every other channel (see the Supplemental Material \cite{suppl3} for the animated data). With further increasing the applied current, vortices near the sample edge are displaced from their equilibrium positions and, therefore, moving vortices are observed in all the channels (see the Supplemental Material \cite{suppl4}). At high temperatures, the pinning strength of the holes reduces, therefore no ordering in the moving vortex lattice is preserved (not shown here). As a consequence the critical current enhancement vanishes with increasing temperature. Thus, the increase of the critical current near half matching field is directly related to the square ice ordering in the vortex structure.

In summary, we studied transport properties of MoGe thin films containing paired nanoscale holes to mimic the elongated double-well pinning sites proposed for the realization of artificial square ice of vortices.\cite{libal} We observed matching effect with striking features never reported before: at half matching field the critical current can be larger and the dissipation can be less than those at first matching field or even at zero-field. Numerical simulations within the Ginzburg-Landau theory showed that the system has a square vortex-ice ground state. With applying external drive, the vortex configuration slightly changes, but still preserves partial ice rules. Simulations also confirmed that the enhancement of the critical current at and around the half matching field is directly related to the square ice vortex arrangements. Although an experimental verification using techniques such as magnetic force microscopy (MFM) imaging is desired, our results indicate that true ice rule obeying systems can be realized with vortices in superconductors, opening an entirely new way to create artificial ice systems. This work also demonstrates that vortex ice can provide an opportunity to probe transport properties that are inaccessible in other types of artificial ice systems.

This work was supported by the US Department of Energy DOE BES under Contract No. DE-AC02-06CH11357 (transport measurements), the Flemish Science Foundation (FWO-Vl) and the Methusalem Foundation of the Flemish Government (numerical simulations). G. R. B. acknowledges individual grant from FWO-Vl. The nanopatterning and morphological analysis were performed at Argonne's Center for Nanoscale Materials (CNM) which is funded by DOE BES under Contract No. DE-AC02-06CH11357. We are grateful to Dr. Charles Reichhardt in Los Alamos National Laboratory for stimulating discussions and critical comments. Z.L.X. acknowledges DOE BES Grant No. DE-FG02-06ER46334 (sample fabrication and imaging). M.L.L was a recipient of the NIU/ANL Distinguished Graduate Fellowship.

\end{document}